\documentclass[twocolumn]{revtex4-2}
\usepackage[utf8]{inputenc}
\usepackage[english]{babel}
\pdfoutput=1 
\usepackage{amsmath}
\usepackage{subcaption}
\usepackage{esint}
\usepackage{ulem}
\usepackage{pgfplots}
\usepackage[margin=2.5 cm]{geometry}
\pgfplotsset{
    compat=newest
}
\usepackage{physics}
\usepackage{graphicx}
\usepackage{mathtools}
\usepackage{wrapfig}
\usepackage{tikz}
\usepackage{etoolbox}
\usetikzlibrary{shapes.geometric}
\usepackage{hyperref}

\begin{document}

\title{Gapless Edge-Modes and Topology in the Qi-Wu-Zhang Model: \\ \large{A Real-Space Analysis}}

\author{Arjo Dasgupta, Indra Dasgupta}

\affiliation{School of Physical Sciences, \\ Indian Association for the Cultivation of Science \\ 2A \& 2B, Raja Subodh Chandra Mallick Rd \\ Jadavpur, Kolkata, West Bengal 700032}

\begin{abstract}
The topological phase transition in the Qi-Wu-Zhang model is studied using a real-space approach. An effective Hamiltonian for the topologically protected edge-modes in a finite-size system is developed. The topological phase transition is understood in terms of a global perturbation to the system which lifts the degeneracy of the edge-modes. The effective Hamiltonian method is also applied to a one-dimensional system with spatially varying hopping strengths to understand the impact of disorder on the edge-modes. 

\end{abstract}

\maketitle

\section{Introduction} 
Chern insulators are two-dimensional materials which are characterised by a topological nontriviality, i.e., a nonzero Chern number \cite{P3},\cite{P6} calculated over the Brillouin Zone. 

Several signatures characterising topologically nontrivial phases have been identified in the literature. The most well-known of these is the appearance of a quantised Hall conductance as in Haldane's original model for a Hall effect without Landau levels \cite{P10}. The obstruction to constructing localised degrees of freedom in nontrivial insulators has recently been exploited \cite{P17} to use the scaling properties of the localisation length to diagnose a topologically nontrivial phase. 

Our aim in this paper is to explain the appearance of chiral edge-modes in the nontrivial phase of a 2D Chern insulator, namely the Qi-Wu-Zhang(QWZ) model \cite{P2}, and develop a method for studying the topological phase transition in this model in terms of the edge-modes alone. 

In order to gain insight into the physical implications of the topology of the system, we carry out a real-space analysis of the system placed on open boundary conditions. This gives us access to details about the differences in the structure of the wavefunctions in the topologically trivial and nontrivial phases. 

We identify a limit within the nontrivial phase of the QWZ model in which all eigenstates for a finite-size system can be determined in closed-form. We show that, through a variable transformation, this limit can be mapped to the familiar nontrivial flat-band limit of the Su-Schrieffer-Heeger Model \cite{P6},\cite{P11}. Considering the deviations from this exactly-solvable limit perturbatively, we construct an effective edge-state Hamiltonian which allows us to study the topological phase transition in terms of the edge-modes alone. We note that the two edge-modes are gapless throughout the nontrivial phase and that opening the gap requires a perturbation acting along the length of the chain, which changes the global structure of the system.

Our analysis of the QWZ model thus brings to light the significance of the Chern number {\it without any reference to the Brillouin Zone}, and thus may be extended to systems lacking lattice translation symmetry. By considering an example with spatially varying hopping strengths in one dimension, we make use of the effective edge-state Hamiltonian to demonstrate the robustness of the edge-modes against disorder.  
 
This paper is organised as follows. The bulk properties of the QWZ model are reviewed in section \ref{qwzmodel}. The properties of the nontrivial dimerised limit are illustrated in section \ref{chrl} and this limit is also compared to the trivial atomic limit. A characterisation of the topological phase transition in terms of the edge-mode degeneracy is illustrated in section \ref{esd}. Finally, the impact of disorder on the edge-modes is considered by studying a modified one-dimensional system with spatially varying hopping strengths in section \ref{dis}. Conclusions are presented in section \ref{conc}.

\section{Bulk Properties: A Brief Review}\label{qwzmodel}

The Qi-Wu-Zhang(QWZ) \cite{P2} model for a Chern insulator is the simplest possible model for a topological insulator defined on a square lattice with an internal degree of freedom ($a/b$ orbital) associated with each lattice site.

In order to study the bulk properties of the model, it may be written in momentum space as:

\begin{eqnarray}\label{QWZ}
 H_{QWZ} ({\bf k}) &=& \lambda \sin(k_x a)\sigma_x+\lambda \sin(k_y a)\sigma_y \\ 
 \nonumber
 &+& (1+\lambda \cos (k_x a)+\lambda \cos(k_y a))\sigma_z 
\end{eqnarray}

The Pauli matrices $\sigma$ are in the $(a/b)$ basis.

The dispersion-relation can be read off from eqn. \ref{QWZ} as:

\begin{eqnarray}\label{disp}
 E_{\pm}({\bf k})&=&\pm [\lambda^2 \sin^2(k_xa)+\lambda^2 \sin^2(k_ya) \\ 
 \nonumber
 &+&(1+\lambda \cos(k_xa)+\lambda \cos(k_ya))^2)]^\frac{1}{2} 
\end{eqnarray}

We note that the gap closes at two specific values of the parameter $\lambda$:

\begin{enumerate}
 \item $\lambda=\frac{1}{2}$: the gap closes at $(k_x,k_y)=(\pm \frac{\pi}{a},\pm \frac{\pi}{a})$.
 \item $\lambda=-\frac{1}{2}$: the gap closes at $(k_x,k_y)=(0,0)$.
\end{enumerate}

shown in Fig. 1(a) and 1(b) respectively.
Near these points, the dispersion relation is approximately linear: they are called Dirac points.

By decomposing $H({\bf k})$ in terms of the Pauli matrices $(\sigma_x,\sigma_y,\sigma_z)$ as ${\bf d}({\bf k}).\vec \sigma$ we find that the parameters trace out a torus in the three-dimensional parameter space given by:

\begin{eqnarray}
\nonumber
  d_x({\bf k})&=&\lambda \sin(k_xa)\\
  d_y({\bf k})&=&\lambda \sin(k_ya)\\
  \nonumber
  d_z({\bf k})&=&1+\lambda \cos(k_xa)+\lambda \cos(k_ya)
\end{eqnarray}

The topology of the system is quantified by the Chern number $n$ which tells us whether the torus encloses the point of degeneracy at the origin. 

For $|\lambda|<\frac{1}{2}$, the parameter $d_z$ is either always positive or always negative. The torus thus misses the origin entirely and the system is topologically trivial (Chern number $n=0$).

For $|\lambda|>\frac{1}{2}$, the system is topologically nontrivial: For $\lambda>\frac{1}{2}$, $n=1$ and for $\lambda<-\frac{1}{2}$, $n=-1$ \cite{P6}.

The Chern number is a topological invariant which captures information about the global structure of the model as we shall see in section \ref{chrl}. Besides, the Kubo formula relates the Hall conductance of the system to the Chern number as: $\sigma_{xy}=\frac{e^2}{2\pi \hbar} n$. A detailed discussion of this topological quantisation of the Hall conductance may be found in \cite{P8}.

\begin{figure}[h]
 \centering
 \includegraphics[width=0.9\columnwidth]{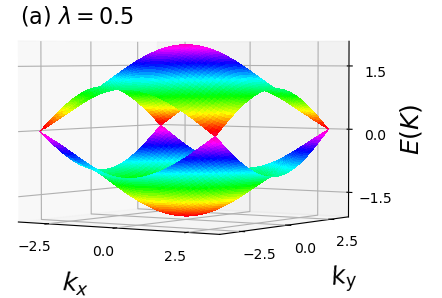}
 \includegraphics[width=0.9\columnwidth]{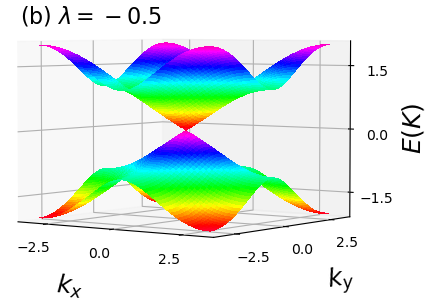}
 \caption{\small Gapless dispersion relations for the QWZ model: Dirac cones appear at (a) $(k_x,k_y) = (\pm \frac{\pi}{a},\pm \frac{\pi}{a})$ for $\lambda=0.5$ and (b) $(k_x,k_y)=(0,0)$ for $\lambda=-0.5$ }
\end{figure}

Thus a topological phase transition takes place between $n=0$ and $n=1$ at $\lambda^+_c = \frac{1}{2}$, and another between $n=0$ and $n=-1$ at $\lambda^-_c=-\frac{1}{2}$. This coincides with the gap-closing at $|\lambda|=\frac{1}{2}$, which makes it impossible to connect the two phases adiabatically \cite{P6}.

\section{Dimensional Reduction and Exactly Localised Edge-Modes}\label{chrl}

The bulk Hamiltonian (see eqn. \ref{QWZ}), whose properties we have discussed in the previous section, assumes periodic boundary conditions and does not tell us about the physics at the edges of the system. To probe the difference in the physical properties of the two topologically distinct phases, it is necessary to study the effects at the edges of the system.

\begin{figure*}[htp]
\includegraphics[width=\columnwidth]{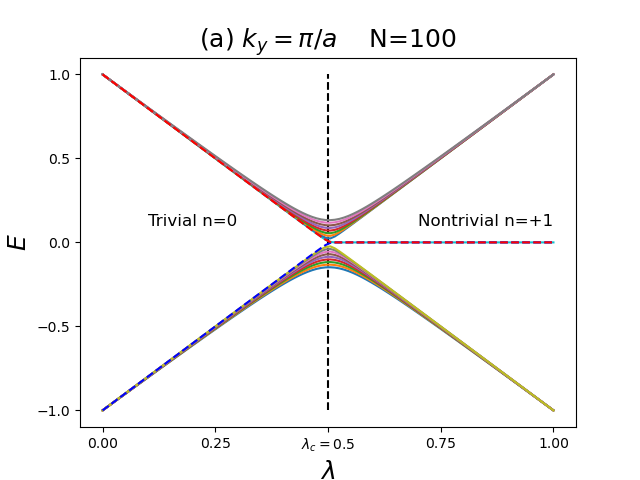}
\includegraphics[width=\columnwidth]{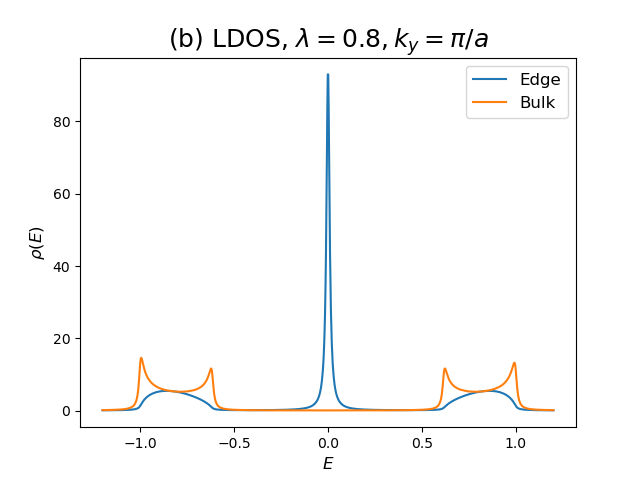}
\caption{\small (a) Low-lying energy levels for $H_{1D}(k_y=\frac{\pi}{a})$  on a lattice of 100 sites: zero-energy edge modes (shown as dashed lines) appear for $\lambda>\frac{1}{2}$. (b) Edge-modes lead to a peak at $E=0$ in the local density of states at the edge.}
\end{figure*}

In order to study the formation of edge-modes, we apply periodic boundary conditions in the $y$-direction and open boundary conditions in the $x$-direction. This allows us to take a Fourier transform along the $y$-direction and separate the full Hamiltonian $H_{QWZ}$ into Hamiltonians describing one-dimensional lattices, each for a  different value of $k_y$. This decomposition into one-dimensional chains is called dimensional reduction \cite{P15}.

\begin{equation}
 H_{QWZ} = \sum_{k_y} H_{1D}(k_y)
\end{equation}

$H_{1D}$ is defined on a 1D chain with an internal degree of freedom ($a/b$ orbital) associated with each lattice site. 

\begin{equation}\label{redHam}
 H_{1D} = \lambda(H_0 + H_1 + V(\lambda))
\end{equation}

Here $H_0$ is a hopping term which breaks time-reversal symmetry:

\begin{eqnarray*}
 H_0 &=& \sum_{i=1}^{N-1} \frac{1}{2}[(a^\dagger_{i+1}a_i - b^\dagger_{i+1}b_i + hc)\\
&+&(ia^\dagger_{i+1}b_i +ib^\dagger_{i+1}a_i + hc)]
\end{eqnarray*}

The breaking of time-reversal symmetry is a necessary condition for the full two-dimensional Hamiltonian to have a non-zero Chern number \cite{P23}. 
 
$H_1$ mixes $a$ and $b$ orbitals at the same lattice site:

\begin{equation*}
 H_1 = \sum_{i=1}^{N} i\sin(k_ya)(b^\dagger_ia_i - a^\dagger_ib_i)
\end{equation*}
 
While $V(\lambda)$ is an onsite term given by:
\begin{equation*}
 V(\lambda) = \sum_{i=1}^{N} \left(\frac{1}{\lambda} + \cos(k_ya)\right)(a^\dagger_ia_i - b^\dagger_ib_i)
\end{equation*}

Note that the operators of the form $a^\dagger_i$ contain a $k_y$ dependence which we have suppressed. $a^\dagger_{i,k_y}$ acting on the vacuum state creates a mode with wavenumber $k_y$ propagating along the $y$-direction localised on the orbital $a$ at the site $i$ on the x-axis. 

As the Hamiltonian $H_{QWZ}$ is quadratic, modes with different values of $k_y$ are independent of each other, leaving us with a family of decoupled one-dimensional chains parameterised by $k_y$. 

By placing the one-dimensional Hamiltonian $H_{1D}$ on a finite lattice (100 sites), we have solved the eigenvalue problem numerically, and demonstrated the formation of edge modes for $\lambda>\lambda_c$ (see Fig. 2). 

As predicted, the edge-modes appear as $\lambda$ crosses the critical value $\lambda_c=\frac{1}{2}$ (see Fig. 2(a)). As $\lambda$ is increased further and the bulk gap opens, the edge-modes remain gapless. 

The difference between the properties of the bulk and edge-states is captured in the local density of states (LDOS)\cite{P20}:

\begin{equation}
  \rho_i(E)=\sum_n |\bra{i}\ket{n}|^2 \delta(E-E_n)
\end{equation}

Where the index $n$ labels the energy-levels and $i$ labels the lattice-sites. 

For $\lambda>\frac{1}{2}$, the localised edge-modes give rise to a peak at $E=0$ in the local density of states at the edges. No such peak appears in the bulk (see Fig. 2(b)).

To study the formation of low-energy edge-modes we look near the $k_y = \frac{\pi}{a}$ Dirac point by setting $k_y=\frac{\pi}{a} + q_y$ for $q_y << \frac{\pi}{a}$, with the system placed in the nontrivial phase ($\lambda>\frac{1}{2}$). 

We start out from the $\lambda=1$ limit, which, as we shall demonstrate, is analytically tractable and provides a good starting point for a perturbative analysis of the entire nontrivial phase.

Considering only terms to first order in $q_y a$, the onsite term $V$ drops out for $\lambda=1$. At this order, the prefactor in $H_1$ is given by $-iq_ya$.

Consider the state $\ket{R} = \frac{1}{\sqrt{2}}(\ket{a_N}-i\ket{b_N})$ localised at the right edge of the system. This state is created by the action of the operator $\Psi_R^\dagger = a^\dagger_N - i b^\dagger_N$ on the vacuum state. We find that $\Psi_R^\dagger$ commutes with $H_0$: 

\begin{eqnarray}
\nonumber
 [H_0,\Psi_R^\dagger] &=& [a^\dagger_{N-1}a_N, a^\dagger_N] + i [b^\dagger_{N-1}b_N, b^\dagger_N] \\
 &-&  i[b^\dagger_{N-1}a_N, a^\dagger_N] - [a^\dagger_{N-1}b_N, b^\dagger_N] \\
 \nonumber
 &=& (a^\dagger_{N-1}-ib^\dagger_{N-1})(b^\dagger_N b_N-a^\dagger_N a_N)\\
 \nonumber
 &+&(a^\dagger_{N-1}-ib^\dagger_{N-1})(a^\dagger_N a_N-b^\dagger_N b_N) = 0
\end{eqnarray}

$\ket{R}$ is thus a zero-energy eigenstate of $H_0$. $\ket{R}$ is also an eigenstate of $H_1$:

\begin{equation}
 H_1\ket{R}= q_ya \ket{R}
\end{equation}

We thus have $H_{1D}\ket{R} =  q_ya \ket{R}$. Similarly, we have the eigenstate $\ket{L}=\frac{1}{\sqrt{2}}(\ket{a_1}+i\ket{b_1})$ localised at the left edge of the system: $H_{1D}\ket{L} = -q_ya \ket{L}$. 

On considering the group velocity $v_y = \frac{\partial E}{\partial q_y}$, we find that the left and right edge-modes propagate in opposite directions. They are thus called chiral edge-modes. 

\vspace{0.5cm}

We find upon deviating from the Dirac point $(k_y=\frac{\pi}{a})$, that the chiral edge-modes are non-degenerate and the one-dimensional system is thus gapped. Hence, to analyse the wavefunctions of the gapless edge-modes, we shall consider the $(\lambda=1, k_y=\frac{\pi}{a})$ limit in which only the term $H_0$ survives. Motivated by the structure of these edge-modes, we define the new variables $\ket{\psi_i} = \frac{1}{\sqrt{2}} (\ket{a_i} - i\ket{b_i})$ and $\ket{\bar{\psi_i}} = \frac{1}{\sqrt{2}} (\ket{a_i} + i\ket{b_i})$.

Under the action of $H_0$, the states $\ket{\psi_i}$ and $\ket{\bar\psi_{i}}$ transform as:

\begin{eqnarray*}
 H_0 \ket{\psi_i} &=& \ket{\bar\psi_{i+1}}\\
 H_0 \ket{\bar\psi_{i}} &=& \ket{\psi_{i-1}}
\end{eqnarray*}

The states $\ket{\psi_i}$ move to the right under the action of $H_0$, while the states $\ket{\bar\psi_i}$ move to the left. 

The states $\ket{R}=\ket{\psi_N}$ and $\ket{L}=\ket{\bar\psi_1}$ at the two edges of the chain thus have nowhere to go and are annihilated by $H_0$, i.e., they are zero-energy edge-modes of $H_0$. These edge-modes correspond to the unpaired Majorana modes which appear at the edges of the one-dimensional model introduced by Kitaev \cite{pa1}.

The presence of the edge-modes make the system gapless even as the bulk of the system is insulating. 

What happens in the bulk? $\ket{\psi_i}$ and $\ket{\bar\psi_{i+1}}$ combine to form $2N-2$ dimerised eigenstates (see Fig. 3(a), $\ket{\alpha_i} = \frac{1}{\sqrt{2}}(\ket{\psi_i}+\ket{\bar\psi_{i+1}})$ and $\ket{\beta_i} = \frac{1}{\sqrt{2}}(\ket{\psi_i}-\ket{\bar\psi_{i+1}})$:

\begin{eqnarray}
 \nonumber
H_0 \ket{\alpha_i} &=& \ket{\alpha_i} \\
H_0 \ket{\beta_i} &=& -\ket{\beta_i}
\end{eqnarray}

The states $\ket{\psi_i},\ket{\bar\psi_i}$  are created by the action of the operators $\Psi^\dagger_i = \frac{1}{\sqrt{2}} (a^\dagger_i - ib^\dagger_i)$ and $\bar\Psi^\dagger_i = \frac{1}{\sqrt{2}} (a^\dagger_i + ib^\dagger_i)$on the vacuum state. It can be shown that these operators follow the same fermionic algebra as the $a_i, b_i$ operators. In terms of these new fermions, the Hamiltonian $H_0$ appears as:

\begin{equation}
 H_0 = \sum_{i=1}^{N-1} (\Psi^\dagger_i \bar\Psi_{i+1} + hc)
\end{equation}

In these new variables, $H_0$ is identical to the nontrivial flat-band limit of the Su-Schrieffer-Heeger (SSH) model for polyacetylene introduced in \cite{P11}. The edge-modes of the Qi-Wu-Zhang model correspond to the instantons in doped polyacetylene which lead to a marked increase in the electrical conductivity of the polymer \cite{P21}. These instantons have also been experimentally observed in optical lattices \cite{P22}. 

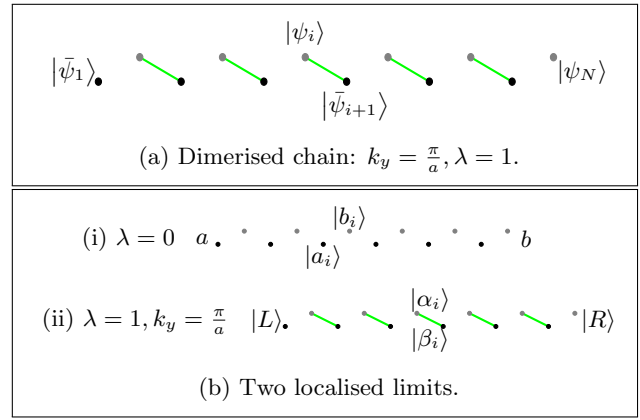
\begin{figure}[h]
\centering
\fbox{
\begin{subfigure}{\columnwidth}
\begin{tikzpicture}[xscale=0.55,yscale=0.65]
  \draw [thick,green] (-1,-2) -- (0,-2.5);
  \draw [thick,green] (1,-2) -- (2,-2.5);
  \draw [thick,green] (3,-2) -- (4,-2.5);
  \draw [thick,green] (5,-2)--(6,-2.5);
  \draw [thick,green] (7,-2)--(8,-2.5);
  \node at (3,-1.5) {$\ket{\psi_i}$};
  \node at (4.2,-3) {$\ket{\bar\psi_{i+1}}$};
  \node at (-2.6,-2.3) {$\ket{\bar\psi_1}$};
  \node at (9.7,-2.3) {$\ket{\psi_N}$};
  \draw [fill,black](-2,-2.5) circle (0.075cm);
  \draw [fill,black](0,-2.5) circle (0.075cm);
  \draw [fill,black](2,-2.5) circle (0.075cm);
  \draw [fill,black](4,-2.5) circle (0.075cm);
  \draw [fill,black](6,-2.5) circle (0.075cm);
  \draw [fill,gray](-1,-2) circle (0.075cm);
  \draw [fill,gray](1,-2) circle (0.075cm);
  \draw [fill,gray](3,-2) circle (0.075cm);
  \draw [fill,gray](5,-2) circle (0.075cm);
  \draw [fill,gray](7,-2) circle (0.075cm);
  \draw [fill,black](8,-2.5) circle (0.075cm);
  \draw [fill,gray](9,-2) circle (0.075cm);
 \end{tikzpicture}
 \caption{\small Dimerised chain: $k_y=\frac{\pi}{a}, \lambda=1$.}
\end{subfigure}}
\fbox{
\begin{subfigure}{\columnwidth}
 \begin{tikzpicture}[scale=0.35]
   [xscale=0.55,yscales=0.65]
  \node at (3,-1.5) {$\ket{b_i}$};
  \node at (2,-3) {$\ket{a_i}$};
  \node at (-2.6,-2.3) {$a$};
  \node at (9.7,-2.3) {$b$};
  
  \draw [fill,black](-2,-2.5) circle (0.075cm);
  \draw [fill,black](0,-2.5) circle (0.075cm);
  \draw [fill,black](2,-2.5) circle (0.075cm);
  \draw [fill,black](4,-2.5) circle (0.075cm);
  \draw [fill,black](6,-2.5) circle (0.075cm);
  \draw [fill,gray](-1,-2) circle (0.075cm);
  \draw [fill,gray](1,-2) circle (0.075cm);
  \draw [fill,gray](3,-2) circle (0.075cm);
  \draw [fill,gray](5,-2) circle (0.075cm);
  \draw [fill,gray](7,-2) circle (0.075cm);
  \draw [fill,black](8,-2.5) circle (0.075cm);
  \draw [fill,gray](9,-2) circle (0.075cm);
  \node at (-5.4,-2.3) {(i) $\lambda = 0$};
 \end{tikzpicture}
 \newline
  \begin{tikzpicture}[scale=0.35]
  [xscale=0.55,yscale=0.65]
  \draw [thick,green] (-1,-2) -- (0,-2.5);
  \draw [thick,green] (1,-2) -- (2,-2.5);
  \draw [thick,green] (3,-2) -- (4,-2.5);
  \draw [thick,green] (5,-2)--(6,-2.5);
  \draw [thick,green] (7,-2)--(8,-2.5);
  \node at (3.5,-1.5) {$\ket{\alpha_i}$};
  \node at (3.5,-3) {$\ket{\beta_i}$};
  \node at (-2.7,-2.3) {$\ket{L}$};
  \node at (9.9,-2.3) {$\ket{R}$};
  
  \draw [fill,black](-2,-2.5) circle (0.075cm);
  \draw [fill,black](0,-2.5) circle (0.075cm);
  \draw [fill,black](2,-2.5) circle (0.075cm);
  \draw [fill,black](4,-2.5) circle (0.075cm);
  \draw [fill,black](6,-2.5) circle (0.075cm);
  \draw [fill,gray](-1,-2) circle (0.075cm);
  \draw [fill,gray](1,-2) circle (0.075cm);
  \draw [fill,gray](3,-2) circle (0.075cm);
  \draw [fill,gray](5,-2) circle (0.075cm);
  \draw [fill,gray](7,-2) circle (0.075cm);
  \draw [fill,black](8,-2.5) circle (0.075cm);
  \draw [fill,gray](9,-2) circle (0.075cm);
  \node at (-7.7,-2.3) {(ii) $\lambda = 1, k_y = \frac{\pi}{a}$};
 \end{tikzpicture}
\caption{\small Two localised limits.}
\end{subfigure}}
\caption{\small (a) The states $\psi_i$ and $\bar\psi_{i+1}$ combine to form dimers, leaving two free ends. (b) The system hosts localised eigenstates at two limits: (i) The atomic limit at $\lambda=0$, and (ii) $\lambda=1, k_y = \frac{\pi}{a}$.}
\end{figure}

The correspondence between the dimensionally-reduced QWZ model ($H_{1D}$) and the SSH model can be made more explicit by looking at the two flat-band limits of $H_{1D}$. Both in the case of the atomic limit  ($\lambda=0$, $H_{1D} = \sum_{i=1}^{N} a^\dagger_ia_i - b^\dagger_ib_i$) and the nontrivial dimerised limit $  (\lambda=1, k_y=\frac{\pi}{a})$  ($H_{1D}=H_0$), the bands are flat, and the states are localised. Despite this, these states have drastically different properties. In the $\lambda=0$ limit the states are restricted to individual atoms $a_i$ and $b_i$ (see Fig. 3(b)(i)). The system preserves time-reversal symmetry and is topologically trivial. On the other hand, the dimerised flat-band limit $(\lambda=1, k_y=\frac{\pi}{a})$ is topologically nontrivial. Despite breaking time-reversal symmetry, the states are localised as a consequence of the dimerisation  of the states $\ket{\psi_{i}}$ and $\ket{\bar\psi_{i+1}}$ (see Fig. 3(b)(ii)). This dimerisation leaves room for zero-energy states at the two edges, making the system gapless.

\vspace{0.5cm}

Since the nearest-neighbour hopping, present in $H_0$, always changes the sublattice index, there are no diagonal terms in the sublattice basis. $H_0$ thus has chiral symmetry\cite{P12}:  $\{H_0,\sigma_z\}=0$, where $\sigma_z$ can be written in terms of the projectors onto the $\psi$, $\bar\psi$ sublattices as $\sigma_z = P_{\psi} - P_{\bar\psi}$. As discussed in \cite{P6}, chiral symmetry has the consequence of restricting all zero-energy eigenstates to a single sublattice -  for instance, $H_0 P_{\psi}\ket{E_n=0} = \frac{1}{2} H_0  (1-\sigma_z) \ket{E_n=0} = 0$. This is precisely the case for the gapless edge-states $\ket{L}$ and $\ket{R}$ in our model.

\vspace{0.5cm}

$H_1$ however breaks the chiral symmetry of the system as can be seen explicitly when we express it in terms of the new variables $\psi_i, \bar\psi_i$: $H_1$ appears as an onsite potential $H_1 = \sum_{i=1}^N \sin(k_ya)({\bar\Psi^\dagger_i\bar\Psi_i -  \Psi^\dagger_i\Psi_i})$. This term does not change the sublattice index and thus breaks chiral symmetry. $H_1$ lifts the degeneracy of the edge-states, and leads to a mixing between the chiral partners $\ket{\alpha_i}$ and $\ket{\beta_i}$ in the bulk.

\vspace{0.5cm}

We note here that a similar analysis can be carried out in the $n=-1$ phase, by expanding about the $k_y=0$ Dirac point and choosing $\lambda=-1$.

\section{Gaplessness and the Topological Phase Transition}\label{esd}

The eigenstates of the dimerised flat-band limit ($\lambda=1, k_y=\frac{\pi}{a}$) are all localised and provide us a useful basis for studying the physics of the QWZ model within the topologically nontrivial phase. In this section, we carry out a perturbative analysis within the subspace of the two edge-states on deviating from the $\lambda=1$ limit. 

We note that any deviation from the Dirac point wavenumber $k_y = \frac{\pi}{a}$ breaks the chiral symmetry of the system and immediately lifts the edge-state degeneracy - the system is thus just a trivial insulator with no gapless states. We thus restrict ourselves to $k_y = \frac{\pi}{a}$ as we tune $\lambda$.

As we decrease $\lambda$ from $\lambda=1$ towards the critical point at $\lambda_c = \frac{1}{2}$, the onsite term $V$ comes into play. The states $\ket{\psi_i}$ and $\ket{\bar\psi_i}$ are connected by the action of $V$: $V\ket{\psi_i} = \xi \ket{\bar\psi_i}$ with $\xi := \frac{1}{\lambda}-1$. Within the nontrivial phase $\lambda>\frac{1}{2}$ and thus $\xi<1$. In the trivial phase $\xi>1$. In effect, V acts as a hopping term for the localised states and thus leads to the delocalisation of both the bulk and the edge-states (see Fig. 4(a)).

\begin{figure}[h]
\centering
\fbox{\begin{subfigure}{\columnwidth}
 \begin{tikzpicture}[xscale=0.55,yscale=0.65]
  \draw [thick,green] (-1,-2) -- (0,-2.5);
  \draw [thick,green] (1,-2) -- (2,-2.5);
  \draw [thick,green] (3,-2) -- (4,-2.5);
  \draw [thick,green] (5,-2)--(6,-2.5);
  \draw [thick,green] (7,-2)--(8,-2.5);
  \draw [dashed,blue] (-2,-2.5)--(-1,-2);
  \draw [dashed,blue] (0,-2.5)--(1,-2);
  \draw [dashed,blue] (2,-2.5)--(3,-2);
  \draw [dashed,blue] (4,-2.5)--(5,-2);
  \draw [dashed,blue] (6,-2.5)--(7,-2);
  \draw [dashed,blue] (8,-2.5)--(9,-2);
  \node at (2,-3) {$\ket{\bar\psi_{i}}$};
  \node at (5.5,-1.5) {$\ket{\psi_{i+1}}$};
  \node at (3,-1.5) {$\ket{\psi_i}$};
  \node at (4.2,-3) {$\ket{\bar\psi_{i+1}}$};
  \node at (-2.6,-2.3) {$\ket{\bar\psi_1}$};
  \node at (9.9,-2.3) {$\ket{\psi_N}$};
  \node at (4.2,-1.95) {$V$};
  \draw [fill,black](-2,-2.5) circle (0.075cm);
  \draw [fill,black](0,-2.5) circle (0.075cm);
  \draw [fill,black](2,-2.5) circle (0.075cm);
  \draw [fill,black](4,-2.5) circle (0.075cm);
  \draw [fill,black](6,-2.5) circle (0.075cm);
  \draw [fill,gray](-1,-2) circle (0.075cm);
  \draw [fill,gray](1,-2) circle (0.075cm);
  \draw [fill,gray](3,-2) circle (0.075cm);
  \draw [fill,gray](5,-2) circle (0.075cm);
  \draw [fill,gray](7,-2) circle (0.075cm);
  \draw [fill,black](8,-2.5) circle (0.075cm);
  \draw [fill,gray](9,-2) circle (0.075cm);
 \end{tikzpicture}
 \caption{Delocalisation of the localised states.}
 \end{subfigure}}
 \fbox{
\begin{subfigure}{0.9864\columnwidth}
  \begin{tikzpicture}
  [xscale=0.55,yscale=0.65]
  \draw [thick,green] (-1,-2) -- (0,-2.5);
  \draw [thick,green] (1,-2) -- (2,-2.5);
  \draw [thick,green] (3,-2) -- (4,-2.5);
  \draw [thick,green] (5,-2)--(6,-2.5);
  \draw [thick,green] (7,-2)--(8,-2.5);
  
  \node at (-2.7,-2.3) {$\ket{L}$};
  \node at (9.9,-2.3) {$\ket{R}$};
  
  \draw [fill,black](-2,-2.5) circle (0.075cm);
  \draw [fill,black](0,-2.5) circle (0.075cm);
  \draw [fill,black](2,-2.5) circle (0.075cm);
  \draw [fill,black](4,-2.5) circle (0.075cm);
  \draw [fill,black](6,-2.5) circle (0.075cm);
  \draw [fill,gray](-1,-2) circle (0.075cm);
  \draw [fill,gray](1,-2) circle (0.075cm);
  \draw [fill,gray](3,-2) circle (0.075cm);
  \draw [fill,gray](5,-2) circle (0.075cm);
  \draw [fill,gray](7,-2) circle (0.075cm);
  \draw [fill,black](8,-2.5) circle (0.075cm);
  \draw [fill,gray](9,-2) circle (0.075cm);
  \node at (7.5,-1.5) {$\ket{\alpha_{N-1}}$};
  \node at (7.5,-3) {$\ket{\beta_{N-1}}$};
  \node at (5.5,-1.5) {$\ket{\alpha_{N-2}}$};
  \node at (5.5,-3) {$\ket{\beta_{N-2}}$};
  \node at (-.45,-1.5) {$\ket{\alpha_1}$};
  \node at (-.45,-3) {$\ket{\beta_1}$};
  \draw [gray][stealth-,thick] (6.3,-1.5)--(6.6,-1.5);
  \draw [blue][stealth-,thick] (6.3,-3)--(6.6,-3);
  \draw [gray][stealth-,thick] (3.3,-1.5)--(4.3,-1.5);
  \draw [blue][stealth-,thick] (3.3,-3)--(4.3,-3);
  \draw [gray][stealth-,thick] (2,-1.5)--(3,-1.5);
  \draw [blue][stealth-,thick] (2,-3)--(3,-3);
  \draw [gray][stealth-,thick] (0.5,-1.5)--(1.5,-1.5);
  \draw [blue][stealth-,thick] (0.5,-3)--(1.5,-3);
  \draw  [gray][stealth-,thick] (8.5,-1.5)--(9,-2);
  \draw  [blue][stealth-,thick] (8.5,-3)--(9,-2);
  \draw  [gray][stealth-,thick] (-2,-2.5)--(-1.3,-1.5);
  \draw  [blue][stealth-] (-2,-2.5)--(-1.3,-3);
\end{tikzpicture}
\caption{Perturbative analysis of the edge-states.}
 \end{subfigure}}
 \caption{\small (a) The localised eigenstates of $H_0$ delocalise under the action of the onsite term $V$. (b) Intermediate states for shortest path between $\ket{R}$ and $\ket{L}$. Hopping from the states $\ket{\alpha}$ to the states $\ket\beta$ is not allowed, drastically reducing the number of possible paths.}
\end{figure}
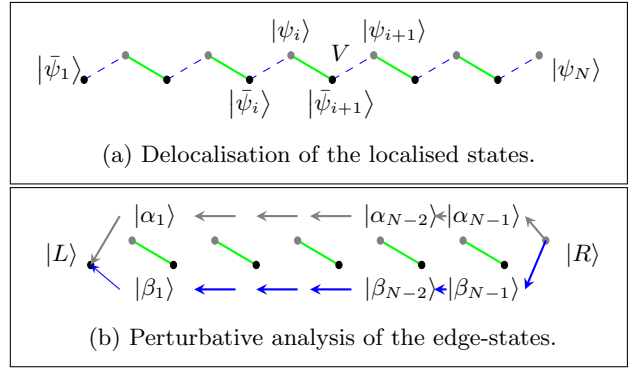

\vspace{0.5cm}
In order for the edge-state $\ket{R}$ to mix with the other edge-state $\ket{L}$, it has to hop across the entire length of the chain. For a lattice of $N$ sites, the degeneracy of the two states is broken by $V$ through a process of order $N$. Below this order, the edge-modes remain gapless.

In order to study the effect of $V$ at this order we make use of Brillouin-Wigner perturbation theory, as discussed in \cite{P5}, to construct an effective Hamiltonian in the edge-state subspace, taking into account the effects of mixing with the bulk-states. The underlying assumption is that the edge-state and bulk-state subspaces are well-separated in energy allowing us to restrict our analysis to the edge-state subspace.

An eigenstate $\ket{n}$ of the unperturbed Hamiltonian $H_0$, receives a correction $\ket{n^{(m)}}$ at order $m$ given by:

\begin{eqnarray}\label{BW}
 \ket{n^{(m)}} &=&  \sum_{\{\zeta_m\}} \ket{\zeta_m} \frac{1}{E_N-\epsilon_{\zeta_m}}\bra{\zeta_m}V\ket{\zeta_{m-1}} \\
 \nonumber
  && \>\>\>\>\>\>\>\>\>\>\>\>\>\> \cross...\cross \ket{\zeta_1}\frac{1}{E_N-\epsilon_{\zeta_1}}\bra{\zeta_1}V\ket{n}
\end{eqnarray}

Here $E_N$ denotes the energy corresponding to the eigenstate $\ket{N}$ of the full Hamiltonian $H_0+V$. The sum is carried out over the intermediate states $\{\ket{\zeta_m}\}$ which are nondegenerate with the unperturbed state $\ket{n}$. 

In our case, we start out from the unperturbed eigenstate $\ket{R}$ of the system. To reach the other end $\ket{L}$ of the chain in $N$ steps, there are 2 possible sets of intermediate states: $\{\alpha_{N-1},...,\alpha_1\}$ or $\{\beta_{N-1},...,\beta_1\}$ (see Fig. 4(b)). This is because all matrix elements of the form $\bra{\alpha_i}V\ket{\beta_j}$ vanish: $\bra{\alpha_i}V\ket{\beta_j} = (\bra{\psi_i}+\bra{\bar\psi_{i+1}})(\xi\ket{\bar\psi_{j+1}}-\xi\ket{\psi_j})=0$.

As there are no energy corrections to $\ket{R}$ up to order $N-1$ in $V$, we can replace the perturbed energy $E_N$ in the energy denominators of the expression in eqn. \ref{BW} with the unperturbed energy $\epsilon_R = 0$.

The $N-1$ order wavefunction correction is thus given by:

\begin{eqnarray}\label{bwig}
\nonumber
 \ket{R^{(N-1)}} &=&  \ket{\alpha_1} \frac{1}{0-1}\bra{\alpha_1}V\ket{\alpha_2} \\
 \nonumber
 && \cross...\cross
 \ket{\alpha_{N-1}}\frac{1}{0-1}\bra{\alpha_{N-1}}V\ket{R}\\
 &+& \ket{\beta_1} \frac{1}{0+1}\bra{\beta_1}V\ket{\beta_2} \\
 \nonumber
 && \cross...\cross
 \ket{\beta_{N-1}}\frac{1}{0+1}\bra{\beta_{N-1}}V\ket{R}
\end{eqnarray}

Other contributions to the wavefunction at this order do not reach the first lattice site and thus do not contribute to the edge-state effective Hamiltonian.

Now, $\bra{\alpha_i}{V}\ket{\alpha_{i+1}}= \xi$ and $\bra{\beta_i}{V}\ket{\beta_{i+1}} = -\xi$.

Thus we obtain the energy correction at order $N$ as:

\begin{equation}\label{edgap}
 E_R^{(N)} = -2\xi^N  
\end{equation}

Reintroducing the factor of $\lambda$ in equation \ref{redHam}, we obtain the $N$-th order effective Hamiltonian for the edge-states in the $\{\ket{R},\ket{L}\}$ basis as: 

\begin{equation}
 V_{edge} = \lambda \begin{bmatrix}
             0 & -2\xi^N\\
             -2\xi^N & 0
            \end{bmatrix}
\end{equation}

\begin{figure}[h]
 \includegraphics[width=1.1\columnwidth]{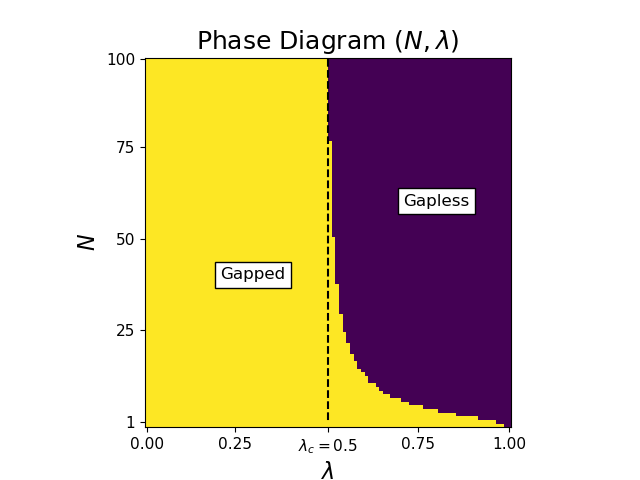}
 \caption{\small Phase diagram from the edge-states for varying values of $\lambda$ and the system size $N$. For small $N$ we find that the edge-state degeneracy is lifted on deviating slightly from $\lambda=1$.}
\end{figure}

There is thus a gap of $\Delta = 4\lambda\xi^N$ between the edge-modes at the lowest nonvanishing order of perturbation theory. We note that the effective Hamiltonian and the result for the energy gap are only valid within the nontrivial phase and the energy gap diverges exponentially in the trivial phase as $N\rightarrow \infty$. This is as expected: the edge-modes no longer exist in the trivial phase, and the perturbation theory breaks down. 

In the nontrivial phase, $\lambda>\frac{1}{2}$ and thus $\xi<1$. The process required to break the degeneracy is suppressed exponentially. This explains why the edge-modes remain gapless throughout the nontrivial phase.

As indicated in eqn. \ref{edgap}, the suppression of the process which breaks the edge-state degeneracy in the nontrivial phase is dependent on the number of links on the lattice. Fig. 5 shows the phase diagram for varying $\lambda$ and system size $N$. We find that for smaller systems, the edge-state degeneracy is lifted even on small deviations from $\lambda=1$ (we consider the edge-states to be gapless if their energy-gap $\Delta_E<0.001$).

The edge-states of the system in the $(\lambda=1,k_y=\frac{\pi}{a})$ limit are confined only to one sublattice: $\ket{R}=\ket{\psi_N}$ and $\ket{L}=\ket{\bar\psi_1}$. On deviating from $\lambda=1$, all the eigenstates of the system have weights on both sublattices, indicating that there are no states exactly at zero energy, as per the argument presented in section \ref{chrl}. For large system-sizes, the energy-gap between the edge-states remains small as long as the system is within the nontrivial phase ($\lambda>\frac{1}{2}$). However, for smaller lattices, this gap becomes more pronounced, as indicated by the phase diagram.

\vspace{0.5cm}

Breaking the degeneracy of the edge-states requires a global perturbation acting along the entire length of the lattice. This is indicative of the fact that this degeneracy is topologically protected and hence robust against local perturbations. 

\section{Effect of Disorder}\label{dis}

The interplay of topology and disorder leads to interesting consequences for two-dimensional Chern insulators. These systems have been studied numerically using real-space versions of the Chern number to characterise the topology of disordered systems as detailed in \cite{pa2} and \cite{pa3}. The Bott index is a useful topological invariant for the numerical study of topological properties in disordered systems, as in the case of the system analysed in \cite{pa4}.
\vspace{0.5cm}

As we have discussed in the previous section, our method for the topological characterisation of the system makes use of the edge-state degeneracy in a one-dimensional chain. Since our treatment of the one-dimensional equivalent of the QWZ model presented in section \ref{esd} does not rely on lattice-translation symmetry, we can apply it to study the topological properties of disordered systems in one dimension.

We shall restrict ourselves to disordered systems in which chiral symmetry is preserved. As we have seen in section \ref{chrl}, perturbations of the form $\sum_{i=1}^N ({\bar\Psi^\dagger_i\bar\Psi_i -  \Psi^\dagger_i\Psi_i})$, which break chiral symmetry, immediately open up a gap between the edge-states.

We consider a modification of the model discussed in the previous section by introducing a spatial variation in the hopping strengths $t_i$ between the unit cells $i$ and $i+1$. $H=\lambda(H_0+V)$, where:

\begin{eqnarray}
H_0 &=& \sum_{i=1}^{N-1} \frac{t_i}{2}[(a^\dagger_{i+1}a_i - b^\dagger_{i+1}b_i + hc)\\
\nonumber
&+&(ia^\dagger_{i+1}b_i +ib^\dagger_{i+1}a_i + hc)]
\end{eqnarray}

and $V$ is given by:

\begin{equation}
 V=\sum_{i=1}^{N} \left(\frac{1}{\lambda}-1\right)(a^\dagger_ia_i-b^\dagger_ib_i)
\end{equation}

As before, in the $\lambda=1$ limit, $V$ drops out. The states $\ket{\psi_n}$ and $\ket{\bar \psi_{n+1}}$ mix under the action of $H_0$:
\begin{equation*}
 H_0\ket{\psi_n} = t_i \ket{\bar \psi_{n+1}} \>\>\>\>\>\>  H_0\ket{\bar \psi_{n+1}} = t_i \ket{\psi_n}
\end{equation*}

These states thus combine to form dimerised eigenstates $\ket{\alpha_i} = \frac{1}{\sqrt{2}}(\ket{\psi_i}+\ket{\bar \psi_{i+1}})$ and $\ket{\beta_i} = \frac{1}{\sqrt{2}}(\ket{\psi_i}-\ket{\bar \psi_{i+1}})$ of $H_0$:

\begin{equation}
 H_0\ket{\alpha_i} = t_i \ket{\alpha_i} \>\>\>\>\>\>  H_0\ket{\beta_i} = -t_i \ket{\beta_i} 
\end{equation}

We take note of the fact that the bulk and edge-subspaces are well-separated in energy only if $t_i\sim 1$ on all sites. The perturbative analysis fails if $t_i<<1$ on some sites.

The edge-states $\ket{R}=\ket{\psi_N}$ and $\ket{L}=\ket{\bar\psi_1}$ remain zero-energy eigenstates of $H_0$. Under the action of $V$, these edge-states mix at order $N$ as before. However, the energy denominators in eqn. \ref{bwig} are no longer constant but vary with the lattice site:

\begin{eqnarray*}
 \ket{R^{(N-1)}} &=&  \ket{\alpha_1} \frac{1}{0-t_1}\bra{\alpha_1}V\ket{\alpha_2} \\
 && ...
 \ket{\alpha_{N-1}}\frac{1}{0-t_{N-1}}\bra{\alpha_{N-1}}V\ket{R}\\
 &+& \ket{\beta_1} \frac{1}{0+t_1}\bra{\beta_1}V\ket{\beta_2} \\
 && ...
 \ket{\beta_{N-1}}\frac{1}{0+t_{N-1}}\bra{\beta_{N-1}}V\ket{R}\\
\end{eqnarray*}

The edge-state effective Hamiltonian changes to:

\begin{equation}
 V_{edge} = \frac{\lambda}{\Pi_{i=1}^{N-1} \> t_i}  \begin{bmatrix}
             0 & -2\xi^N\\
             -2\xi^N & 0
            \end{bmatrix}
\end{equation}

In order to prevent the edge-state degeneracy from being lifted, the geometric mean $t$ of the hopping strengths $t_i$, must be smaller than $\xi$. The critical point $\lambda_c$ is thus given by:

\begin{equation}
   \frac{1}{\lambda_c}-1 = \xi_c= (\Pi_{i=1}^{N-1}  \> t_i)^\frac{1}{N} =: t
\end{equation}

In order to compare this result to the numerical solution, we generate sets of hopping strengths $t_i$ drawn from a uniform random distribution centered around a mean value $\bar t $ with a spread of $0.1$ in each case. The mean $\bar t$ is chosen such that the values of $t_i$ are not negligible compared to the bulk band-gap, which would potentially lead to the formation of new edges and alter the physics drastically (In our case we have chosen $4$ values of $\bar t$, between $0.50$ and $1.35$). For each set of hopping strengths $\{t_i\}$, we evaluate the corresponding values of the geometric mean $t$, and diagonalise the Hamiltonian numerically. We plot the energy-gap $\Delta_E$ between the two edge-states as a function of the parameter $\lambda$ to detect the critical point, at which $\Delta_E$ reaches $0$ (we consider the system to be gapless if $\Delta_E<0.001$). The values of $\xi_c = \frac{1}{\lambda_c}-1$ thus obtained are found to be in close agreement with the predicted value $\xi_c=t$ (see Fig. 6).

\begin{figure}[h]
 \includegraphics[width=\columnwidth]{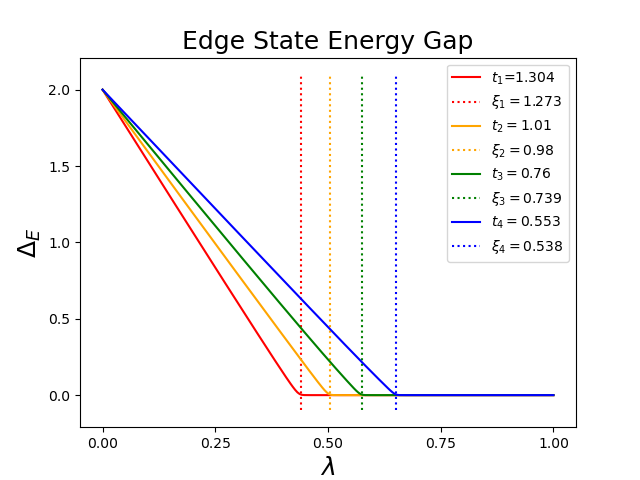}
 \caption{\small Gap between edge states for disordered chains. The geometric mean $t$ of the hopping strengths determines the critical point.}
\end{figure}

Note that the critical point is insensitive to the details of the configuration $\{t_i\}$ of hopping strengths, and depends only on the geometric mean $t$. This demonstrates the robustness of the edge-states against disorder.

As we have noted, the perturbative argument breaks down for configurations in which $t_i<<1$ on some sites. This corresponds to the problem of bond-dilution as studied in \cite{P16} and has more interesting consequences for the edge-states.

\section{Conclusion}\label{conc}

The real-space analysis of the Qi-Wu-Zhang model we have presented in sections \ref{chrl} and \ref{esd}  makes use of the exactly-solvable limit within the nontrivial phase to obtain an exact expression for the edge-state wavefunctions. Deviations from this limit allow us to understand the topological phase transition in terms of the {\it delocalisation and ultimate disappearance} of these gapless edge-modes as the system passes from the nontrivial to the trivial phase. We note that only a {\it global perturbation} can break the edge-mode degeneracy and change the topological class of the system.

The real-space description of the dimensionally-reduced QWZ model we have discussed does not rely on lattice-translation symmetry. The edge-state effective Hamiltonian thus allows us to demonstrate the robustness of the edge-states of the one-dimensional system against disorder, as we have demonstrated in section \ref{dis}.

Our analysis of topological properties in the presence of disorder, which is limited only to one-dimensional chiral-symmetric systems, may be compared to the analysis in \cite{pa7}, where the control of edge-states in a one-dimensional superconducting circuit lattice is discussed. It is shown that introducing disorder (implemented using spatially-varying hopping strengths, as in section \ref{dis}) causes local fluctuations in the bulk states but does not affect the edge-states. This is consistent with our findings in section \ref{dis} and demonstrates the robustness of the edge-states in a one-dimensional model.

By making a direct contact between the topological phase and the edge-state degeneracy, the effective Hamiltonian approach we have employed gives us an intuitive picture of the topological phase transition, both in the case of the clean two-dimensional QWZ model and the disordered version of the equivalent one-dimensional model.

Our analysis is significant in that it both illustrates the formation of edge-modes and their topological protection without any reference to the reciprocal space and illustrates the connection between the edge-state degeneracy and the topology of the system directly.
\vspace{0.5 cm} 

\paragraph*{\bf Acknowledgements} We would like to thank Ritwik Das and Pritam Sarkar for useful discussions and Sudeshna Dasgupta for valuable comments on the manuscript. 

\paragraph*{\bf Declaration} The authors have no conflicts of interest to declare.

Note: All figures, except graphs, have been have been created using the tikz package. \url{https://tikz.dev/}
\end{document}